\documentclass[aps,pre,reprint,twocolumn,superscriptaddress,showpacs,floatfix,longbibliography]{revtex4-2}
\usepackage[latin1]{inputenc}
\usepackage{epsfig}
\usepackage{amsfonts}
\usepackage{amssymb}
\usepackage{calrsfs}
\usepackage{amsmath}
\usepackage{xcolor}
\usepackage{graphicx}
\usepackage[colorlinks=true,allcolors=blue]{hyperref}
\hypersetup{breaklinks=true}

\begin{document}
	
	\title{Oscillating states of driven Langevin systems in large viscous regime}
	\author{Shakul Awasthi}
	\email{shakulawasthi010615@iisertvm.ac.in}
	\author{Sreedhar B. Dutta}
	\email{sbdutta@iisertvm.ac.in}
	\affiliation{School of Physics, Indian Institute of Science Education and Research Thiruvananthapuram, Thiruvananthapuram 695551, India}
	
	\date{\today}
	
	\begin{abstract} 
		We employ an appropriate perturbative scheme in the large viscous regime to study oscillating states in driven Langevin systems. We explicitly determine oscillating state distribution of under-damped Brownian particle subjected to thermal, viscous and potential drives to linear order in anharmonic perturbation. We also evaluate various non-equilibrium observables relevant to characterize the oscillating states. We find that the effects of viscous drive on oscillating states are measurable even in the leading order and show that the thermodynamic properties of the system in these states are immensely distinct from those in equilibrium. 
	\end{abstract}

	\maketitle
	
	\newcommand{\bee}{\begin{equation}}
		\newcommand{\eee}{\end{equation}}
	\newcommand{\tm}{(t)}
	\newcommand{\gm}{\gamma}
	\newcommand{\dbar}{d\hspace*{-0.08em}\bar{}\hspace*{0.1em}}
	\newcommand{\haf}{\frac{1}{2}}
	\newcommand{\x}{\mathbf{x}}
	
	%
	%
	%
	%
	
\section{Introduction}\label{intro}
Many-particle systems can exist in a variety of states exhibiting drastically different behavior when subjected to appropriate macroscopic conditions. There has been an increasing interest to investigate the behavior of various physical observables in a variety of systems under periodically driven conditions\,\cite{Jung1993,Brandner2015,Dutta2003,Dutta2004,Wang2015,Knoch2019,Kim2010,Fiore2019,Koyuk2018,Oberreiter2019,Busiello2018,Holubec2020,Koyuk2019,Datta2021}. Periodic drives can induce macroscopic systems to be in an {\it oscillating state} which is a stable non-equilibrium state with time-dependent and periodic thermodynamic properties. 

Unlike equilibrium or steady states\,\cite{Zhang2012-1,Seifert2010,Seifert2012,Qian2006}, very little is known about the nature of the oscillating states. In general, the thermodynamic properties of the oscillating states cannot be deduced by any perturbative phenomenological modifications of equilibrium properties. Presumably studying the appropriate stochastic dynamics of certain relevant macroscopic quantities of driven systems and their asymptotic distributions may provide insights toward understanding of these states.  While it may not be easy to establish the conditions under which a generic system can exist in oscillating states, recent studies\,\cite{Awasthi2020,Awasthi2021} show that the driven Langevin systems, even under anharmonic perturbations, can exist and persist in these states. Both the conditions of existence and the probability distributions of the oscillating states were in fact established by treating the periodic drive exactly to all orders of perturbation in anharmonicity. However the dependence on the periodic parameters of the driven system is established only implicitly. Hence even though, in principle, one can obtain the expectation of various relevant observables in the oscillating states of the Langevin system, it is hard to decipher any underlying thermodynamic relations that may exist between them. 
This motivates us to explore the possibility of extracting explicitly the dependence of observables on the drive parameters. 

A driven Langevin system is specified by various $T$-periodic parameters namely, the viscous strength$~\gamma$, the noise strength$~D$ and those associated with the potential$ ~U$. In absence of driving under certain conditions the Langevin system asymptotically reaches an equilibrium state whose distribution depends only on the bath temperature$~{T_b=D/\gamma}$ and the potential$~U$. For a fixed temperature, the equilibrium state has no memory of the viscous coefficient$~\gamma$. On the other hand, in the presence of driving under appropriate conditions the system asymptotically reaches an oscillating state that carries the viscous memory unlike an equilibrium state. Therefore, in order to parameterize the state space of the oscillating states of this system, not only do we need to extend the state space variables of the corresponding equilibrium states to $T$-periodic functions of time, but also need to include additional nonequilibrium variables such as$~\gamma$. Essentially for a fixed temperature drive various viscous drives retain the system in different oscillating states. This naturally motivates us to address the question, what is the dependence of oscillating state on viscous drive? To this end, we explore the state space with a one parameter extension of viscous drive$~\gamma_{\alpha}(t)= \alpha \gamma(t)$ for any given positive periodic functions$~\gamma(t)$ and$~T_b(t)$, where$~\alpha$ is a positive real parameter. More precisely, in order to understand the nature of oscillating states we will evaluate various relevant observables for large values of$~\alpha$ where we have a good perturbative control and determine their viscous dependence. In the process we will also address the question, what are the characteristic features of these states that make them distinct from equilibrium?

In the next section, we introduce a class of driven under-damped anharmonic Langevin systems, familiarize with essential features and properties of its oscillating states and then specify some relevant observables. In Sec.\,\ref{OS-large-al}, we develop a perturbative scheme applicable in the large viscous limit and determine explicitly the probability distribution of oscillating states. In Sec.\,\ref{ther-obs}, we obtain various thermodynamic observables to sub-leading order in$~1/\alpha$ and investigate their dependence on thermal and viscous drives. In Sec.\,\ref{ther-obs}, we briefly verify numerically a few interesting and elusive features of oscillating states, and finally in Sec.\,\ref{con-rem}, we summarize and conclude with some remarks.

\section{Oscillating states of driven Langevin systems}\label{OS-PDSS}

In this section, we will first briefly recollect some essential aspects of the oscillating states of driven Langevin system. We will then write down the probability distribution of the oscillating states in presence of quartic perturbations. Furthermore we will list some of the observables which are relevant for characterizing these nonequilibrium states. 

\subsection{Probability distribution}

A periodically driven underdamped Brownian particle is a prototypical example of a driven Langevin system described by the stochastic variables$~X_t$ and$~V_t$ whose dynamics is governed by the equations, 
	\begin{align}\label{stoc-dyn}
	\dot{X}_t& = V_t ~,\nonumber \\
	\dot{V}_t& = -\gamma V_t +f \left( X_t; k, \lambda \right) + \eta(t)~,
\end{align} 
where$~\gamma, f$ and $\eta$ denote the viscous strength, the external force and the noise, respectively. The force$~f = -\partial_x U(x; k, \lambda)$ can depend both on the harmonic strength$~k$ and any number of anharmonic couplings$~\lambda$. The noise is taken to be Gaussian with zero mean and non-zero variance$~{\langle \eta\tm \eta(t') \rangle_{\eta}= 2D(t) \delta(t-t')}$ of strength$~D$. The effect of periodic driving is accounted for by considering some or all of the parameters$~\gamma, D, k$ and$~\lambda$ to be time dependent with period$~T$. 

The asymptotic state of the driven Langevin system under certain conditions is $T$-periodic and is referred to as an oscillating state. The dependence of these states on the periodic drive can be determined exactly by exploiting the underlying $SL_2$ symmetry\,\cite{Awasthi2020,Awasthi2021}. The probability distribution$~P_{os}^{(0)}(x,v,t)$ of the oscillating states in absence of anharmonic terms in the drive is Gaussian and hence can be specified by its covariance matrix$~\Sigma$ whose matrix elements are$~{\Sigma_{11} =\langle x^2 \rangle_0} $, ${\Sigma_{12}= \Sigma_{21}=\langle xv \rangle_0}$ and$~{\Sigma_{22}=\langle v^2 \rangle_0}$. These second moments in the harmonic case with force$~f=-kx$ are governed by the dynamical equations
\begin{eqnarray}\label{mom-2-as}
	\frac{d}{dt}\Sigma_{11} &=& 2 \Sigma_{12}~, \nonumber \\
	\frac{d}{dt}\Sigma_{12} &=&-k \Sigma_{11} -\gamma \Sigma_{12} + \Sigma_{22}~,\nonumber \\
	\frac{d}{dt}\Sigma_{22} &=& -2k \Sigma_{12} -2 \gamma\Sigma_{22} +2D~,
\end{eqnarray}
and can be obtained from their arbitrary solutions up on imposing $T$-periodicity.

 The underlying $SL_2$ symmetry in the harmonic Langevin dynamics can be exploited\,\cite{Awasthi2020} to write down not only second moments but all moments in terms of the solutions of an associated Hill equation
 \begin{equation}\label{hill-eq-0}
 	\frac{d^2u}{dt^2}  + \nu u =0~,\quad \nu = k  - \frac{1}{2}\dot{\gamma} - \frac{1}{4}\gamma^2~.
 \end{equation}
More importantly, a pseudoperiodic choice of solutions$~{u}_{\pm}$ to the Hill equation, ${u}_{\pm}(t+T) ={u}_{\pm}(t) \exp{(\mu_{\pm}T)}$, enables us to read the Floquet exponents$~{\mu_{\pm}=\pm \mu}$ that are required to establish the existence of oscillating states. Essentially, the necessary condition for the existence and stability of the oscillating states,
 \begin{equation}\label{cond-mu-g0}
 	| Re{(\mu)} | < \frac{1}{2} \overline{\gamma}~,
 \end{equation} 
depends on$~\mu$. 

 Astonishingly the anharmonic perturbations do not alter the criterion\,\eqref{cond-mu-g0} for stability\,\cite{Awasthi2021} but of course will introduce corrections to the distribution. We will restrict in this work to quartic anharmonic case with potential
 \begin{equation}
 U=\frac{1}{2}kx^2 + \epsilon\frac{1}{4}\lambda x^4
 \end{equation}
 where $\epsilon$ is a book-keeping parameter of the order of perturbation. Furthermore we will limit the calculations to first order even though the analysis can be straightforwardly extended to more general cases and/or higher orders. In this case the oscillating state distribution to $O(\epsilon)$ is given by
\begin{equation}\label{P1}
	P_{os}(x,v,t) = \left[1 -\epsilon \left( A^{(1)} - \langle A^{(1)} \rangle_0 \right)\right]P_{os}^{(0)}(x,v,t)~,
\end{equation}
where $\langle A^{(1)} \rangle_0$ is average of $A^{(1)}$ with respect to $P_{os}^{(0)}$, 
\begin{equation}\label{A1}
	A^{(1)} = \sum_{r=0}^2 \tilde{a}_r x^{2-r} v^r+ \sum_{r=0}^4 a_r x^{4-r} v^r~,
\end{equation}
and the $T$-periodic coefficients$~\tilde{a}_r$ and$~a_r$ satisfy certain dynamical equations which are established\,\cite{Awasthi2021} by substituting Eq.\eqref{P1} in the corresponding Fokker-Planck equation of the driven Langevin system\eqref{stoc-dyn} and then equating the coefficients of independent monomials to zero. 
These dynamical equations for the quartic potential reduce to 
\begin{equation} \label{a4eom}
\frac{d}{dt}	
\begin{bmatrix}
	a_0\\ a_1\\ a_2\\a_3\\a_4
\end{bmatrix}= 
	\begin{bmatrix}
		0 & k_p & 0 & 0 & 0\\
		-4 & \gamma_p & 2 k_p&0 & 0\\
		0 & -3 &2\gamma_p&3k_p & 0\\
		0 & 0 & -2 & 3\gamma_p & 4 k_p\\
		0 & 0 &0 &-1 & 4\gamma_p
	\end{bmatrix} 
\begin{bmatrix}
	a_0\\ a_1\\ a_2\\a_3\\a_4
\end{bmatrix}
+ \lambda
\begin{bmatrix}	
	\Sigma^{-1}_{12}\\ \Sigma^{-1}_{22}\\0\\0\\0
\end{bmatrix},
\end{equation}
and
\begin{equation} \label{a2eom}
	\frac{d}{dt}	
	\begin{bmatrix}
		\tilde{a}_0\\ \tilde{a}_1\\ \tilde{a}_2
	\end{bmatrix}= 
	\begin{bmatrix}
		0 & k_p & 0 \\
		-2 & \gamma_p & 2 k_p\\
		0 & -1 &2\gamma_p
	\end{bmatrix} 
	\begin{bmatrix}
		\tilde{a}_0\\ \tilde{a}_1\\ \tilde{a}_2	 
	\end{bmatrix}
	+ 2D
	\begin{bmatrix}
		a_2\\3a_3\\6a_4
	\end{bmatrix},
\end{equation}
where
\begin{equation}\label{gkp}
	\gamma_p := \gamma-2D\Sigma^{-1}_{22} ,\quad k_p := k - 2 D \Sigma^{-1}_{12} 
\end{equation}
and$~\Sigma$ is the covariance matrix of the asymptotic distribution$~P_{os}^{(0)}(x,v,t)$ of the driven system in the absence of anharmonic terms. Note that the notation$~\Sigma^{-1}_{ij}$ is used for$~(\Sigma^{-1})_{ij}$ for any$~i$ and$~j$. The initial conditions of both the dynamical equations are fixed by demanding $T$-periodicity of their respective solutions. 

We make a few digressive remarks that are extremely useful in obtaining the moments and distribution when anharmonic terms contain higher-degree polynomials and/or are evaluated to higher-order perturbative corrections.
The solutions of Eqs.\eqref{mom-2-as},\eqref{a4eom} and\,\eqref{a2eom} can be written in the form
\begin{equation}\label{floq-soln}
	\mathbf{a}(t)
	= \kappa(t,0) \mathbf{a}(0)
	+ \int_{0}^{t} ds \kappa(t,s)\mathbf{d}(s)~,
\end{equation}
where $\kappa(t,s) = {\cal M}(t){\cal M}^{-1}(s)$ matrices can be obtained from the fundamental matrices$~{\cal M}$ of Eqs.\eqref{mom-2-as},\eqref{a4eom} and\,\eqref{a2eom}, respectively, and the vector$~\mathbf{d}$ denotes the the corresponding inhomogeneous terms. The underlying$~SL_2$ symmetry permits us to express the fundamental matrix of Eq.\eqref{mom-2-as} completely in terms of the solutions of Hill equation\,\eqref{hill-eq-0}. Similarly, the fundamental matrices of Eqs.\eqref{a4eom} and\,\eqref{a2eom} can be expressed in terms of the solutions of another associated Hill equation 
\begin{equation}\label{hill-eq-1}
	\frac{d^2}{dt^2} u_p + \nu_p u_p  =0~,\quad \nu_p = k_p  - \frac{1}{2}\dot{\gamma}_p - \frac{1}{4}\gamma_p^2~.
\end{equation}
The algorithm for systematically constructing fundamental matrices for any anharmonic perturbation in terms of the solutions of this Hill equation\,\eqref{hill-eq-1} is detailed in Ref.\cite{Awasthi2021}. The effectiveness of this algorithm becomes increasingly evident at higher-order corrections of the distribution where corresponding fundamental matrices become increasingly larger. In this work though where we have restricted to high viscous regimes and first-order perturbation in quartic anharmonicity we solve the systems of differential equations by direct methods as the required calculations are not too cumbersome.  

To summarize, the probability distribution$~P_{os}(x,v,t)$ of the oscillating state of the driven Langevin system that is parameterized by$~k, \gamma, D$ and$~\lambda$ is given by Eq.\eqref{P1} and is completely specified by the $T$-periodic solutions of Eqs.\eqref{mom-2-as},\eqref{a4eom} and\,\eqref{a2eom}. Thus our main objective here is to explicitly specify the harmonic moments$~\{\Sigma_{ij}\}$ and the coefficients$~\{a_i,\tilde{a}_j\}$ in various viscous regimes and then study properties of observables in the oscillating states.

\subsection{Relevant observables}

In the remainder of the section, we will introduce some of the relevant observables of the system in oscillating states and also make apparent that for a given mechanical drive in a thermal environment of temperature$~T_b$ the states can be characterized by the viscous parameter$~\gamma$.

The relevant thermodynamic observables of the oscillating states are not only energy, entropy and other quantities that we encounter in equilibrium but also should include nonequilibrium observables related to rate of work, heat flux, entropy flux, entropy production, and other such quantities that emerge in thermodynamic systems when subjected to driving.
A stochastic thermodynamic description can be employed to define various statistical observables wherein a corresponding stochastic variable$~{g_t=g(X_t,V_t)}$ is associated with each observable$~g$ and a value$~\widetilde{g}(t)$ is ascribed at any time$~t$ given by the relation
\begin{equation}\label{ob-def}
	\widetilde{g}(t) = \langle g(X_t, V_t)\rangle := \int dxdv g(x,v) P_{os}(x,v,t)~.
\end{equation} 
For instance, the stochastic variables associated with energy$~E$\,\cite{Sekimoto1998} and entropy$~S$\,\cite{Seifert2005} are
\begin{eqnarray}
	E_t :=\frac{1}{2} V_t^2 + U(X_t,t)~, \\
	Y_t := -\log P_{os}(X_t,V_t,t)~,
\end{eqnarray}
respectively, where the notation$~U(x,t)$ is used for$~U(x;k(t),\lambda(t))$ for simplicity and the more familiar notations$~E$ and$~S$ are used for energy and entropy instead of$~\widetilde{E}$ and $~\widetilde{Y}$, respectively. 

If the oscillating states can be characterized by a set of thermodynamic variables then any statistical observable has a specific thermodynamic interpretation. How different is the thermodynamic interpretation of the oscillating states from that of equilibrium? To this end, we will investigate various statistical observables including the kinetic energy and the correlation between$~X_t$ and$~V_t$ which in equilibrium have simple and trivial interpretations, respectively. More precisely, we define a quantity$~T_s$ which is twice the average kinetic energy,
\begin{equation}\label{ts-def}
	T_s = \langle V_t^2 \rangle~,
\end{equation}
referred to as temperature of the system or kinetic temperature, and the correlation function
\begin{equation}\label{c-def}
	C = \frac{\langle X_tV_t\rangle}{\sqrt{\langle X_t^2\rangle\langle V_t^2\rangle}}~.
\end{equation}

The stochastic process\eqref{stoc-dyn} also induces dynamics for any stochastic variable$~{g_t=g(X_t,V_t,t)}$. Hence it follows with the Stratonovich interpretation that the first law of thermodynamics$~dE_t = \dbar Q_t + \dbar W_t$ holds strongly\,\cite{Sekimoto1998}, where
\begin{eqnarray}\label{QW-def}
\dbar Q_t &:=&  -\gamma V_t^2 dt + V_t \circ dB_t~, \nonumber \\
 	\dbar W_t &:=& dt\frac{\partial}{\partial t } U(X_t, t)~,
 \end{eqnarray}
 the quantity$~{ dB_{t} =\int_t^{t+dt}dt' \eta(t') }$ denotes the Brownian noise and the symbol$~\circ$ denotes the Stratonovich product. The combination$~{-\gamma V_t + \eta(t)}$ in the Langevin equation\eqref{stoc-dyn} when interpreted as the thermal force naturally leads to the stochastic variable$~\dbar Q_t$ associated with the work done by the heat bath and also provides the identification of$~\langle \dbar Q_t \rangle$ with the infinitesimal heat gained by the system. Similarly the interpretation of$~f \left( X_t; k, \lambda \right)$ as the mechanical force allows the identification of$~\langle dW_t \rangle$ with the infinitesimal work done on the system.
 
 In an oscillating state all the observables of the system are $T$-periodic. Hence the differential functions$~\langle dg_t \rangle$ with Stratonovich convention when integrated over a period vanishes, namely 
 \begin{equation}\label{avg-df}
 	\oint \langle dg_t \rangle := \int_0^T dt \frac{\langle dg_t \rangle}{dt}= \int_0^T dt \frac{d}{dt} \langle g_t \rangle =0~.
 \end{equation}
 Here we have used the fact that$~{\langle dg_t \rangle= d\langle g_t \rangle}$ which is straightforward to prove for any stochastic variables of the form$~{g(X_t,V_t,t)}$ governed by the Langevin dynamics\eqref{stoc-dyn}. Now it is easy to see from the first law that
 \begin{eqnarray}\label{e-rate}
 	\frac{dE}{dt}=	\frac{\langle \dbar W_t \rangle}{dt} + \frac{\langle \dbar Q_t \rangle}{dt}~, 
 \end{eqnarray}
 and in an oscillating state
 \begin{equation}\label{avg-wq}
 	\oint \langle dW_t \rangle  = -\oint \langle \dbar Q_t \rangle~.
 \end{equation} 
 It is of course not unexpected when the system is in an oscillating state that in a period the average work done on it is equal to the heat dissipated by it. It is evident that the value$~\oint \langle \dbar Q_t \rangle$ is completely determined by the oscillating state distribution for given$~\gamma$ and$~D$. In fact the rate of heat dissipation is an observable of the system in an oscillating state and will be referred to as the housekeeping heat flux
 \begin{equation}\label{hk-heat}
 	q_{hk} := -\frac{\langle \dbar Q_t \rangle}{dt} = \gamma \langle V_t^2 \rangle - D. 
 \end{equation}
 The first law also then implies that rate of work$~\langle \dbar W_t \rangle/dt$ is not an independent thermodynamic observable. Unlike in a steady state where the housekeeping heat flux is necessarily a constant, in an oscillating state$~q_{hk}$ is $T$-periodic. 
 
We can also identify some relevant nonequilibrium observables from the rate of change of entropy\,\cite{Seifert2005}
\begin{eqnarray}\label{s-rate}
\frac{dS}{dt}=	\frac{\langle dY_t \rangle}{dt} =  \frac{\gamma}{D} \frac{\langle \dbar Q_t \rangle}{dt} +  \frac{1}{D} \left\langle \left( \frac{J_v^{ir}}{P_{os}} \right)^2_t \right\rangle ~, 
\end{eqnarray}
 where the stochastic variable of the last term is based on the irreversible part of the probability current
\begin{equation}\label{irr-prob-curr}
J_v^{ir} (x,v, t) = - \left( \gamma v  + D \frac {\partial}{\partial v} \right) P_{os}(x,v,t)~.
\end{equation}
The expression\,\eqref{s-rate} follows by simple manipulation\,\cite{Awasthi2020} of the stochastic differential of$~Y_t$ after using Eq.\eqref{stoc-dyn}, and can be rewritten as
\begin{eqnarray}\label{s-r}
	\frac{dS}{dt}=	\Pi-\Phi~, 
\end{eqnarray}
where the rate of entropy production in the system
\begin{eqnarray}\label{ent-pr}
\Pi :=\frac{1}{D} \left\langle \left( \frac{J_v^{ir}}{P_{os}} \right)^2_t \right\rangle~, 
\end{eqnarray}
and the entropy flux from the system to the bath
\begin{eqnarray}\label{ent-fl}
	\Phi := -\frac{1}{T_b} \frac{\langle \dbar Q_t \rangle} {dt} = \gamma \frac{T_s-T_b}{T_b}~,
\end{eqnarray}
provided we identify in accordance with irreversible thermodynamics the temperature of the bath to be
\begin{equation}
	T_b= D/\gamma~.
\end{equation}
This identification is consistent with the fact that oscillating states should reduce to equilibrium states in absence of driving. The entropy flux$~\Phi$ and the entropy production$~\Pi$ are completely determined for given$~\gamma$ and$~T_b$ by the distribution$~P_{os}(x,v,t)$ and indeed are the observables of system in an oscillating state. Since the perpetuation of the system in an oscillating state is necessarily accompanied by the two quantities$~\Phi$ and$~\Pi$, we can refer to them as housekeeping entropy flux and housekeeping entropy production, respectively. Note that these quantities are unlike the entropy flux and production associated with an irreversible process that takes a system from one equilibrium state to another and whose values depend on the specific process. Essentially, the house-keeping flux$~\Phi$ and production$~\Pi$ are genuine non-equilibrium observables that may be required to characterize the oscillating states. It is evident from Eq.\eqref{s-r} that$~\Phi$ and$~\Pi$ are though not independent. In fact we could decide to monitor energy, entropy and any one of the three quantities$~q_{hk}, \Phi$ and$~\Pi$.  
Note though that in absence of mechanical drive 
\begin{eqnarray}\label{p-ind}
	\Pi =\frac{dS}{dt} -\frac{1}{T_b} \frac{dE}{dt} ~. 
\end{eqnarray}

We will now proceed to investigate the thermodynamic behavior of the system in oscillating states for any given bath temperature$~T_b$ as we vary the viscous parameter$~\gamma$.

\section{Probability distribution in large viscous limit}\label{OS-large-al}

In this section, we will determine the probability distributions of the oscillating states for a given thermal drive$~T_b$ along a one-parameter extension$~\gamma_{\alpha} =\alpha \gamma$ of the viscous drive$~\gamma$, where$~\alpha$ is a large dimensionless constant. To this end, we will develop the appropriate perturbation scheme, analyze the existence condition\,\eqref{cond-mu-g0} and evaluate the harmonic moments$~\{\Sigma_{ij}\}$ and the coefficients$~\{a_i,\tilde{a}_j\}$ in ${1/\alpha}$-expansion.

\subsection{Perturbative scheme}

Under the one-parameter extension, the Hill equation\,\eqref{hill-eq-0} gets modified to
	\begin{equation}\label{hill-eq-0-a}
	\frac{d^2u_{\alpha}}{dt^2}  + \nu_{\alpha} u_{\alpha} =0~,\quad \nu_{\alpha} = k  - \frac{1}{2}\alpha\dot{\gamma} - \frac{1}{4}\alpha^2\gamma^2~.
\end{equation}
The solutions of the modified Hill equation can be determined by employing the semi-classical expansion  
\begin{equation}\label{wkb-exp}
	u_{\alpha}= e^{\alpha \varphi}~, \quad \varphi = \sum_{n=0}^{\infty} \frac{1}{\alpha^n} \varphi^{(n)}~.
\end{equation}
where the Taylor coefficients$~\varphi^{(n)}$ of$~\varphi$ satisfy certain differential equations which can obtained by substituting Eq.\eqref{wkb-exp} in Eq.\eqref{hill-eq-0-a} and then equating the coefficients of all the monomials of$~\alpha$ and$~1/\alpha$ to zero. These differential equations are found to be
\begin{eqnarray}\label{phi-eqs}
&&(\dot{\varphi}^{(0)})^2 - \frac{1}{4}\gamma^2 =0~, \nonumber \\
&&\ddot{\varphi}^{(0)}+2	\dot{\varphi}^{(0)}\dot{\varphi}^{(1)} -\frac{1}{2}\dot{\gamma} =0~, \nonumber \\
&&\ddot{\varphi}^{(1)}+2	\dot{\varphi}^{(0)}\dot{\varphi}^{(2)} 
+ (\dot{\varphi}^{(1)})^2  + k=0~, \nonumber \\
&&\ddot{\varphi}^{(n-1)}+\sum_{r=0}^{n}	\dot{\varphi}^{(r)}\dot{\varphi}^{(n-r)} =0~,~\text{for~}n\ge 3~,
\end{eqnarray}
and whose structure clearly suggests that the coefficients$~\dot{\varphi}^{(n)}$ can be determined not only hierarchically but almost algebraically. Note that we are not considering rapidly driven Langevin systems\,\cite{Dutta2003} and have essentially assumed that$~\alpha \gg 1/\gamma T$.

 It is evident that there are two independent solutions$~\varphi_{\pm}$ corresponding to$~\dot{\varphi}^{(0)}_{\pm} = \pm\gamma/2$. Since$~\dot{\varphi}_{\pm}$ are clearly $T$-periodic, the corresponding solutions$~u_{\alpha\pm}$ are indeed the pseudo-periodic solutions of the Hill equation with Floquet exponents
\begin{equation}\label{fexp-h}
	\mu_{\alpha\pm} = \frac{\alpha}{T}\left(\varphi_{\pm}(T) - \varphi_{\pm}(0) \right) = \alpha \overline{\dot{\varphi}_{\pm}}.
\end{equation}  
where we have used over-line notation$~\overline{f}$ to indicate average of a periodic function$~f(t)$ over a time period, namely 
\begin{equation}
  \overline{f} := \frac{1}{T} \int_{0}^{T}\!dt~ f(t)~.
\end{equation} 
It is straightforward to write down the series expansion of$~\dot{\varphi}_{\pm}$ which for instance to sub-sub-leading order is given by
\begin{eqnarray}\label{phidots}
	&&\dot{\varphi}_{+} =\frac{\gamma}{2} - \frac{1}{\alpha^2}\frac{k}{\gamma}  + \cdots, \\
	&&\dot{\varphi}_{-} =-\frac{\gamma}{2} -\frac{1}{\alpha}\frac{\dot{\gamma}}{\gamma} + \frac{1}{\alpha^2}
	\left( \frac{k}{\gamma} -\frac{d}{dt}\frac{\dot{\gamma}}{\gamma^2}\right) + \cdots,
\end{eqnarray}
and in turn leads to the Floquet exponents 
\begin{equation}\label{floq-subsub}
\mu_{\alpha\pm} = \pm\frac{\alpha}{2} \overline{\gamma}\mp \frac{1}{\alpha} \overline{k/\gamma} + \cdots.
\end{equation}
Hence for large$~\alpha$ the existence condition\,\eqref{cond-mu-g0} for the oscillating states will continue to hold provided $\overline{k/\gamma} > 0$, which is in fact a weaker condition than demanding a positive harmonic potential at all times.

\subsection{Taylor coefficients of moments}

When the oscillating state exists for given$~T_b$ and $~{\gamma_{\alpha}= \alpha\gamma}$ for large$~\alpha$ then the $T$-periodic moments have Taylor expansion
\begin{equation}\label{taylor-mom}
\Sigma_{ij} = \sum_{n=0}^{\infty} \frac{1}{\alpha^n}\Sigma_{ij}^{(n)}~,
\end{equation}
where we have suppressed the$~\alpha$ index on$~\Sigma_{ij}$ and the indices$~i$ and$~j$ run over$~1$ and$~2$. The Taylor coefficients satisfy a hierarchy of differential equations obtained from the one-parameter extension of Eqs.\eqref{mom-2-as} where ${\gamma \to \alpha \gamma}$ and ${D \to \alpha \gamma T_b}$. 
Before we discuss these equations and their solutions a few comments on large$~\alpha$ expansion are in order. The extensions of Eqs.\eqref{mom-2-as} in the limit$~\alpha \to \infty$ result in differential equations of lower order and hence the limit is singular. 
We know that when the existence condition of the oscillating state holds then the asymptotic solution$~\Sigma_{ij}(t;\alpha)$ is bounded, periodic and independent of the initial conditions. Essentially, the relaxation times of moments$~\Sigma_{11}, \Sigma_{12}$ and$~\Sigma_{22}$ are $1/\left(\alpha\overline{\gamma}-2|\mu_{\alpha}|\right)$,
$1/\alpha\overline{\gamma}$ and
$1/\left(\alpha\overline{\gamma}+2|\mu_{\alpha}|\right)$, respectively\,\cite{Awasthi2020}.
Further, if the existence condition holds for some$~\alpha$ and in its neighborhood then the asymptotic solution is Taylor expandable. It is imperative that the limit $\alpha \to \infty$ should be taken only after taking the large-time limit. Hence when we solve for the Taylor coefficients from the hierarchy of dynamical equations we impose periodicity of the solutions instead of taking large-time limit of an arbitrary solution.

We will now proceed to obtain the Taylor series iteratively. The leading order equations can be read from $O(\alpha)$ terms of the one-parameter extension of Eqs.\eqref{mom-2-as} and are given by
\begin{equation}\label{zero-eqn}
	\Sigma_{22}^{(0)} = T_b~, \quad \Sigma_{12}^{(0)} = 0~, 
\end{equation}
and
\begin{equation}\label{zero-eqn2}
 \frac{d}{dt}\Sigma_{11}^{(0)} = 2 \Sigma_{12}^{(0)}~,
\end{equation}
which result in the solution
\begin{equation}
	\Sigma_{11}^{(0)} =  c^{(0)}~,
\end{equation}
where$~c^{(0)}$ is a constant. Since we have assumed$~{\alpha \gg 1/\gamma T}$ the leading order equations and the whole set of perturbative hierarchy have to be accordingly modified when rapid drives are considered. 

Suppose$~\Sigma_{ij}^{(n)}$ is determined for a given$~{n \ge 0}$, then we can obtain the next-order coefficients$~\Sigma_{ij}^{(n+1)}$ from $n$-th order equations in the hierarchy given by
\begin{eqnarray}\label{ite-mom}
&&\Sigma_{22}^{(n+1)} =-\frac{1}{2\gamma}	\left( \frac{d}{dt}\Sigma_{22}^{(n)} +2k \Sigma_{12}^{(n)} \right)~, \nonumber \\
&&\Sigma_{12}^{(n+1)} =-\frac{1}{\gamma}	\left( \frac{d}{dt}\Sigma_{12}^{(n)} +k \Sigma_{11}^{(n)} -\Sigma_{22}^{(n)} \right)~,	\nonumber \\
&&\frac{d}{dt}\Sigma_{11}^{(n+1)} = 2 \Sigma_{12}^{(n+1)}~.
\end{eqnarray}
The last two equations lead to the expression 
\begin{equation}\label{sig-11-n}
	\Sigma_{11}^{(n+1)}(t)	= c^{(n+1)} + \widehat{\Sigma}_{11}^{(n+1)}(t)~,
\end{equation}
 where the hatted quantity is introduced for later convenience and is given by 
\begin{equation}\label{sig-11-hat}
\widehat{\Sigma}_{11}^{(n+1)}(t)	= 2\int_{0}^{t} \!\!ds \;\Sigma_{12}^{(n+1)}(s)~,
\end{equation}
Though Eq.\eqref{sig-11-n} involves an integration constant$~c^{(n+1)}$, $\Sigma_{11}^{(n+1)}$ is not arbitrary due to $T$-periodicity of the solution at next order. Essentially note that the average of$~\Sigma_{12}^{(n+2)}$ over a period vanishes and hence the constant of integration$~c^{(n+1)}$ gets fixed by using the next-order condition
 \begin{equation}
\int_{0}^{T} \!\!dt \; \Sigma_{12}^{(n+2)}(t) =0 .	
 \end{equation}
Also note that this condition holds even for$~n=-1,-2$.

 When $O(\alpha^0)$ dynamics is included then we obtain
\begin{eqnarray}\label{1-ord-sig}
&&\Sigma_{22}^{(1)} =-\frac{\dot{T_b}}{2\gamma} ~,~\Sigma_{12}^{(1)} = \frac{1}{\gamma} \left(T_b - c^{(0)}k \right)~, \nonumber \\
&&\Sigma_{11}^{(1)} = c^{(1)} + 2\int_{0}^{t} \!\!ds \; \Sigma_{12}^{(1)}(s) ~,
\end{eqnarray}
where$~c^{(1)}$ is a constant. At this order the constant$~c^{(0)}$ though not$~c^{(1)}$ gets determined as mentioned earlier due to the periodicity of$~\Sigma_{11}^{(1)}$ and is given by the expression
\begin{equation}\label{c0-def}
	c^{(0)}= \frac{\overline{(T_b/\gamma)}}{\overline{(k/\gamma)}}~.
\end{equation} 
We can thus proceed to write down the next order coefficients$~\Sigma_{22}^{(2)}, \Sigma_{12}^{(2)}$ and$~\Sigma_{11}^{(2)}$ using Eqs.\eqref{ite-mom}. We only list the coefficient $\Sigma_{12}^{(2)}$ and $\Sigma_{22}^{(2)}$ that we require later which are given by
\begin{align}
	\Sigma_{12}^{(2)} &= -\frac{1}{\gamma}	\left( \dot{\Sigma}_{12}^{(1)} +k \Sigma_{11}^{(1)} -\Sigma_{22}^{(1)} \right)~,\\
	\Sigma_{22}^{(2)} &= \frac{1}{\gamma} \frac{d}{dt} \left( \frac{\dot{T_b}}{\gamma} \right) - \frac{4 k}{\gamma} \left( T_b - c^{(0)} k \right)~.
\end{align}
At this order we can fix the constant $c^{(1)}$ in Eqs.\eqref{1-ord-sig} by demanding that the time-average of$~\Sigma_{12}^{(2)}$ over a period vanishes, which gives
\begin{equation}
	c^{(1)} = \frac{1}{\overline{(k/\gamma)}}\left[ \overline{(\Sigma_{22}^{(1)}/\gamma)} - \overline{(\dot{\Sigma}_{22}^{(1)}/\gamma)} -  \overline{(k \widehat{\Sigma}_{11}^{(1)}/\gamma)} \right]~.
\end{equation}

The matrix elements of$~\Sigma^{-1}$ specifically$~\Sigma^{-1}_{22}$ and$~\Sigma^{-1}_{12}$ are required for determining anharmonic corrections of the oscillating state distribution. The matrix elements of$~\Sigma^{-1}$ are Taylor expandable  
\begin{equation}\label{taylor-inv}
	\Sigma^{-1}_{ij} = \sum_{n=0}^{\infty} \frac{1}{\alpha^n} \sigma_{ij}^{(n)}~,
\end{equation}
where the coefficients$~\sigma_{ij}^{(n)}$ can be determined from the Taylor expanded covariance matrix. These coefficients to second order are obtained as follows,
\begin{eqnarray}
&&\sigma_{22}^{(0)} = \frac{1}{T_b}~,~\sigma_{12}^{(0)} = 0~,~\sigma_{11}^{(0)} = \frac{1}{c^{(0)}}~, \nonumber \\
&&\sigma_{22}^{(1)} =-\frac{\Sigma_{22}^{(1)}}{T_b^2} ~,~ 
	 \sigma_{12}^{(1)} = -\frac{\Sigma_{12}^{(1)}}{T_b c^{(0)}}~,~ \sigma_{11}^{(1)} =-\frac{\Sigma_{11}^{(1)}}{(c^{(0)})^2} ~,\nonumber \\
&& \sigma_{22}^{(2)} =\frac{1}{T_b^2} \left(-\Sigma_{22}^{(2)} + \frac{(\Sigma_{12}^{(1)})^2}{c^{(0)}} +\frac{(\Sigma_{22}^{(1)})^2}{T_b}\right)~,\nonumber \\
&& \sigma_{12}^{(2)} = \frac{1}{T_b c^{(0)}} \left(-\Sigma_{12}^{(2)} + \frac{\Sigma_{11}^{(1)}  \Sigma_{12}^{(1)}}{c^{(0)}} +\frac{\Sigma_{12}^{(1)} \Sigma_{22}^{(1)}}{T_b}\right)~,\nonumber \\
&& \sigma_{11}^{(2)} = \frac{1}{(c^{(0)})^2} \left(-\Sigma_{11}^{(2)} + 
\frac{(\Sigma_{11}^{(1)})^2}{c^{(0)}} +\frac{(\Sigma_{12}^{(1)})^2}{T_b} \right)~.
\end{eqnarray}

\subsection{Anharmonic coefficients }

It is now evident that the one-parameter extension of the parameters defined in Eq.\eqref{gkp} is of the form
\begin{equation}\label{p-para-extn}
\gamma_p = - \alpha\gamma +\sum_{n=0}^{\infty} \frac{1}{\alpha^n}\gamma^{(n)}~, \quad k_p= \sum_{n=0}^{\infty} \frac{1}{\alpha^n} k^{(n)}~,
\end{equation}  
where the coefficients are given by 
\begin{equation}
	\gamma^{(n)}= - 2 \gamma T_b \sigma_{22}^{(n+1)}~, \quad
	k^{(n)}= k \delta_{n,0} - 2 \gamma T_b \sigma_{12}^{(n+1)}~.
\end{equation}  
In the oscillating state the coefficients$~a_i$ and$~\tilde{a}_j$ are $T$-periodic and have Taylor expansion 
\begin{equation}
	a_i = \sum_{n=0}^{\infty} \frac{1}{\alpha^n}a_i^{(n)}~, \quad \tilde{a}_j= \sum_{n=0}^{\infty} \frac{1}{\alpha^n} \tilde{a}_j^{(n)}~,
\end{equation} 
respectively. The time dependence of the Taylor coefficients can be extracted for large$~\alpha$ from the one-parameter extensions of Eqs.\eqref{a4eom} and\,\eqref{a2eom} perturbatively. Note that $\tilde{a}~$variables are coupled to $a~$variables by the parameter$~D$ whose one-parameter extension$~\alpha \gamma T_b$ is proportional to$~\alpha$. Hence in order to determine$~\tilde{a}_j^{(n)}$ we need to know  $a~$variables to at least $(n+1)$-th order in$~1/\alpha$.

We will now express all the$~a$ and$~\tilde{a}$ variables of the  oscillating state distribution to sub-leading order in$~1/\alpha$. 
From $O(\alpha^1)$ dynamics which can be read from $O(\alpha^1)$ terms of one-parameter extended Eq.\eqref{a4eom} it is easy to see that
\begin{equation}
	a^{(0)}_1 =a^{(0)}_2 =a^{(0)}_3 =a^{(0)}_4 =0~.
\end{equation} 
When $O(\alpha^0)$ dynamics is included then we obtain the expressions
\begin{eqnarray}
&&a^{(0)}_0 = c^{(0)}_0~, \quad	a^{(1)}_1 = -\frac{1}{\gamma} \left( 4c^{(0)}_0  -\frac{\lambda}{T_b}  \right)~, \nonumber \\
&&a^{(1)}_2 =a^{(1)}_3 =a^{(1)}_4 =0~,
\end{eqnarray}
where$~c^{(0)}_0$ is a constant. When $O(\alpha^{-1})$ dynamics is also included then we obtain the equation
\begin{equation}\label{01aeq}
	\frac{d}{dt} a^{(1)}_0  =k^{(0)} a^{(1)}_1 + \lambda \sigma_{12}^{(1)}~,
\end{equation}
which upon integration gives 
\begin{align}\label{01a}
	&a^{(1)}_0  = c^{(1)}_0 + \widehat{a}^{(1)}_0~,\nonumber\\
	&\widehat{a}^{(1)}_0  =  \int_{0}^{t}ds\left[k^{(0)} a^{(1)}_1 + \lambda \sigma_{12}^{(1)}\right](s) ~,
\end{align}
where$~c^{(1)}_0$ is a constant and $\widehat{a}^{(1)}_0$ is defined for later convenience. Taking the time average of Eq.\eqref{01aeq} and using the periodicity of$~a^{(1)}_0$ determines the constant
\begin{equation}
c^{(0)}_0	= \frac{1}{4}\frac {\overline{(\lambda/\gamma)}}{\overline{(T_b/\gamma)}}~.
\end{equation}
Furthermore at this order we obtain the coefficient 
\begin{align}\label{12a22a-1}
	&a^{(2)}_1~=~4 c_0^{(1)}/\gamma~+~\widehat{a}^{(2)}_1~,\nonumber\\
	&\widehat{a}^{(2)}_1 =\frac{1}{\gamma} \left( -\dot{a}^{(1)}_1 -4\widehat{a}^{(1)}_0 +\gamma^{(0)} a^{(1)}_1 + \lambda \sigma_{22}^{(1)} \right)~.
\end{align}
At this order we also get 
\begin{eqnarray}\label{12a22a-2}
&&a^{(2)}_2 =-\frac{3}{2\gamma}a^{(1)}_1~,\quad
	  a^{(2)}_3 =a^{(2)}_4 =0~.
\end{eqnarray} 
Note that the constant$~c^{(1)}_0$ is still not fixed and hence$~a^{(2)}_1$ is only determined up to a constant. In order to fix this we need to use the following relation from $O(\alpha^{-2})$ dynamics, 
\begin{equation}\label{c10-cond}
	\overline{( k^{(0)} a^{(2)}_1 )} + 	\overline{ ( k^{(1)} a^{(1)}_1 )} + \overline{(\lambda \sigma_{12}^{(2)} )} =0~,
\end{equation}
which can be rephrased as
 \begin{equation}\label{c10-const}
 	c^{(1)}_0 = \frac{1}{4\overline{(k^{(0)}/\gamma)}}\left[	\overline{( k^{(0)} \widehat{a}^{(2)}_1 )} + 	\overline{ ( k^{(1)} a^{(1)}_1 )} + \overline{(\lambda \sigma_{12}^{(2)} )}\right].
 \end{equation}

We will also later need the following expression obtained at this order,
\begin{eqnarray}
a^{(3)}_2 =\frac{1}{2\gamma} \left( -\dot{a}^{(2)}_2 -3a^{(2)}_1 + 2 \gamma^{(0)} a^{(2)}_2 \right) ~.
\end{eqnarray} 

We can similarly proceed to calculate the coefficients of$~\tilde{a}_j$ from one-parameter extended Eq.\eqref{a2eom}. It is easy to see that $O(\alpha^1)$ dynamics leads to
\begin{equation}
	\tilde{a}^{(0)}_1 =\tilde{a}^{(0)}_2 = 0~,
\end{equation} 
while $O(\alpha^0)$ dynamics gives
\begin{equation}
	\tilde{a}^{(0)}_0 =\tilde{c}^{(0)}_0 ~,~\tilde{a}^{(1)}_1 =- \frac{2}{\gamma}\tilde{c}^{(0)}_0 ~,~\tilde{a}^{(1)}_2 = 0~,
\end{equation}
where$~\tilde{c}^{(0)}_0$ is a constant. At $O(\alpha^{-1})$ we obtain the equation
\begin{equation}\label{a10-diff-eq}
	\frac{d}{dt} \tilde{a}^{(1)}_0 = k^{(0)} \tilde{a}^{(1)}_1 + 2\gamma T_b a^{(2)}_2~,
\end{equation}
whose average over a period determines the constant
\begin{equation}
	\tilde{c}^{(0)}_0 = 0~,
\end{equation} 
which leads to $\tilde{a}^{(0)}_0 =\tilde{a}^{(1)}_1 =0$. Moreover Eq.\eqref{a10-diff-eq} also leads to the expression
\begin{align}\label{ta10}
	&\tilde{a}^{(1)}_0 =\tilde{c}^{(1)}_0 + \widehat{\tilde{a}}^{(1)}_0~,\nonumber\\
	&\widehat{\tilde{a}}^{(1)}_0 = 	- 3\int_{0}^{t}ds~ T_b(s)a^{(1)}_1(s)~,
\end{align} 
and$~\tilde{c}^{(1)}_0$ is a constant. Furthermore we get
\begin{equation}
	\tilde{a}^{(2)}_1 =-\frac{2}{\gamma}\tilde{a}^{(1)}_0 ~,~
	\tilde{a}^{(2)}_2 =0~.
\end{equation} 
The constant$~\tilde{c}^{(1)}_0$  gets determined at $O(\alpha^{-2})$ from the relation
\begin{eqnarray}
\overline{(  k^{(0)}\tilde{a}^{(2)}_1 )}+ 2\overline{(\gamma T_b a^{(3)}_2 )} =0~,
\end{eqnarray}
which can be expressed as
 \begin{equation}\label{ctil-const}
	\tilde{c}^{(1)}_0= \frac{1}{\overline{(k^{(0)}/\gamma)}}\left[ \overline{(\gamma T_b a^{(3)}_2 )} - \overline{( k^{(0)} \widehat{\tilde{a}}^{(1)}_0 /\gamma )}  \right] ~.
\end{equation} 

To summarize the large viscous perturbative analysis, we have explicitly obtained the harmonic moments$~\{\Sigma_{ij}\}$ and the coefficients$~\{a_i,\tilde{a}_j\}$ to sub-leading order in$~1/\alpha$. We find that $a^{(0)}_0, a^{(1)}_0, a^{(1)}_1$ and$~\tilde{a}^{(1)}_0$ are the only non-zero components of the coefficients at this order.

\section{Thermodynamic observables in oscillating state}\label{ther-obs}

We can now study the dependence of any observable on bath-temperature$~T_b$ and the viscous drive$~\gamma$ once we express it in terms of$~\{\Sigma_{ij}\}$ and$~\{a_i,\tilde{a}_j\}$. We will first express various relevant quantities in suitable forms and then analyze their leading and sub-leading behavior.

\subsection{Expressions up to $O(\lambda)$}

 The expectation of any observable$~g=g(x,v)$ in the oscillating state can be expressed using Eq.\eqref{P1} as
\begin{eqnarray}
\langle g\rangle
 &=&\langle g \rangle_0 - \left[\tilde{a}_0 \langle g :x^2:\rangle_0
 + \tilde{a}_1 \langle g :xv:\rangle_0 + \tilde{a}_2 \langle g :v^2:\rangle_0 \right. \nonumber \\
 &+& a_0  \langle g :x^4:\rangle_0 + a_1 \langle g:x^3 v:\rangle_0 + a_2  \langle g :x^2v^2:\rangle_0 \nonumber \\
 &+& \left. a_3 \langle g:xv^3:\rangle_0 
 + a_4 \langle g:v^4:\rangle_0 \right], 
\end{eqnarray} 
where$~:f:$ denotes$~{f-\langle f \rangle_0} $ for any function$~{f=f(x,v)}$. Furthermore, any quantity$~\langle g:f:\rangle_0$ can be written in terms of moments$~\Sigma_{ij}$ using Wick's decomposition. Since we are interested in determining the observables to$~O(1/\alpha)$ we can of course drop the coefficients $\tilde{a}_1, \tilde{a}_2, a_2, a_3, a_4$ which vanish at this order. But it turns out that to determine some quantities to$~O(1/\alpha)$ we require$~O(1/\alpha^2)$ terms and hence we will not exclude$~\tilde{a}_1$ and$~a_2$ which are non-vanishing coefficients among them at$~O(1/\alpha^2)$. 

It is straightforward though in some cases cumbersome to Wick decompose the moments and other observables. The expressions we obtained for the moments are
\begin{widetext}
\begin{eqnarray}\label{Wick-mom}
	\langle x^2 \rangle &=&  \Sigma_{11} -2\tilde{a}_0\Sigma_{11}^2-2\tilde{a}_1\Sigma_{11} \Sigma_{12}- 12 a_0 \Sigma_{11}^3 - 12 a_1 \Sigma_{11}^2\Sigma_{12} 
	- 2 a_2\Sigma_{11} \left(5\Sigma_{12}^2 + \Sigma_{11}\Sigma_{22}\right) ,
	\nonumber \\
		\langle xv \rangle &=& \Sigma_{12} - 2\tilde{a}_0\Sigma_{11}\Sigma_{12}- \tilde{a}_1\left(\Sigma_{11}\Sigma_{22} +\Sigma_{12}^2 \right)- 12 a_0 \Sigma_{11}^2\Sigma_{12} -  3 a_1 \Sigma_{11}\left(3\Sigma_{12}^2 + \Sigma_{11}\Sigma_{22}\right) - 4a_2\Sigma_{12} \left(2\Sigma_{11}\Sigma_{22} + \Sigma_{12}^2\right) ,	\nonumber \\
	\langle v^2 \rangle &=& \Sigma_{22} -2\tilde{a}_0\Sigma_{12}^2 
	-2\tilde{a}_1\Sigma_{12}\Sigma_{22}
	- 12 a_0 \Sigma_{11}\Sigma_{12}^2 - 6 a_1 \Sigma_{12}\left(\Sigma_{11}\Sigma_{22} + \Sigma_{12}^2\right) - 2 a_2\Sigma_{22} \left(5\Sigma_{12}^2 + \Sigma_{11}\Sigma_{22}\right).
\end{eqnarray}
\end{widetext}
From these expressions we can read off the kinetic temperature$~T_s$, the correlation function$~C$, the housekeeping heat flux$~q_{hk} = \alpha\gamma \left( T_s - T_b\right)$ defined in Eq.\eqref{hk-heat} and the entropy flux$~\Phi= \alpha\gamma \left( T_s - T_b\right)/T_b$ as defined in Eq.\eqref{ent-fl}. The entropy production rate$~\Pi$ on the other hand can be evaluated once it is expressed in a convenient form as
\begin{eqnarray}\label{Wick-ent}
\Pi = \alpha\gamma \left[ \frac{1}{T_b} \langle v^2\rangle -2 +
\left( \frac{1}{|\Sigma|}+2 a_2 \right)T_b\Sigma_{11}\right],
\end{eqnarray}
which is obtained upon substituting Eqs.\eqref{irr-prob-curr} and\,\eqref{P1} in Eq.\eqref{ent-pr} and carrying out straightforward algebraic manipulations. Similar substitutions reduce the irreversible current$~J_v^{ir}$ defined in Eq.\eqref{irr-prob-curr} to the form 
\begin{eqnarray}\label{Jirr-1}
	J_v^{ir}(x,v,t) = \alpha\gamma P_{os} \left[ -v  +\frac{T_b}{|\Sigma|} \left(\Sigma_{11} v - \Sigma_{12}x \right) \right. \nonumber \\
	+ \left. T_b \left(\tilde{a}_1 x + a_1 x^3 +2a_2 x^2 v \right)\right].
\end{eqnarray}

The energy$~E$ can of course be easily read as
\begin{equation}\label{ener-exp}
	E = \frac{1}{2} \langle v^2\rangle + \frac{1}{2} k\langle x^2\rangle + \frac{3}{4} \lambda \Sigma_{11}^2~.
\end{equation}
Entropy on the other hand requires to be cast in a convenient form to analyze. Up on substituting$~P_{os}$ from Eq.\eqref{P1} in the relation$~S=-\int P_{os}\log P_{os}$ and going through similar manipulation we obtain the expression for entropy as
\begin{eqnarray}\label{entr-exp}
	S&=& \log\left(2\pi e\sqrt{|\Sigma|}\right) 
	-\tilde{a}_0\Sigma_{11} -\tilde{a}_1\Sigma_{12}- 6 a_0 \Sigma_{11}^2 \nonumber \\ &-& 6 a_1 \Sigma_{11}\Sigma_{12} 
	- 2 a_2 \left(2\Sigma_{12}^2 + \Sigma_{11}\Sigma_{22}\right).
\end{eqnarray}

\subsection{Viscous influence on oscillating state at leading order}

The thermodynamic properties of the driven anharmonic Langevin system of course depend on viscous drive but we will now emphasize that this is the case even at the leading order. We will see that the oscillating states in the limit$~{\alpha \to \infty}$ are actually distinguishable from equilibrium. Taking this limit in Eq.\eqref{Wick-mom} will reduce the moments to
\begin{eqnarray}\label{mom-inf}
	\langle x^2 \rangle &\to& \langle x^2 \rangle^{(0)} = \Sigma_{11}^{(0)}  - 12 a_0^{(0)}  \Sigma_{11}^{(0)3},
	\nonumber \\
	\langle xv \rangle &\to& 0  ,	\nonumber \\
	\langle v^2 \rangle &\to& \langle v^2 \rangle^{(0)} =\Sigma_{22}^{(0)} = T_b.
\end{eqnarray}
Essentially the leading order probability distribution of the oscillating state has a few similarities with that of equilibrium. For instance, the position and velocity variables are uncorrelated and the kinetic temperature$~T_s$ of the system is same as the bath temperature$~T_b$ though time dependent.   
The dissimilarities can be noticed even for harmonic drives where, for instance, the equipartition of energy no longer holds as the difference$~{T_b - c^{(0)} k}$ is in general non-zero. It is interesting to note that the variance of position$~\langle x^2 \rangle^{(0)}$ at leading order unlike that of velocity is not determined by the instantaneous values of the driving parameters.
 
We notice from Eq.\eqref{mom-inf} that$~\langle v^2 \rangle$ is independent of$~\gamma$ while$~\langle x^2 \rangle$ is not when$~\gamma$ is time dependent. Note that$~\Sigma_{11}^{(0)} (=c^{(0)})$ and$~a_0^{(0)} (=c_0^{(0)})$ are independent of$~\gamma$ either when$~\gamma$ is time independent or when both$~T_b$ and potential parameters$~k$ and$~\lambda$ are time independent. Even in case of viscous drive both$~\Sigma_{11}^{(0)}$ and$~a_0^{(0)}$ are invariant when$~\gamma$ is scaled by a constant. In general it is clear that the distribution and in turn observables in oscillating state depend on viscous drive and the effects can be measured right from leading order.

\subsection{Viscous dependence of observables to sub-leading order}

We now Taylor expand various thermodynamic quantities to$~O(1/\alpha)$ and study their behavior. The kinetic temperature$~T_s$ of the system to$~O(1/\alpha)$ can be easily read from the expression for$~\langle v^2 \rangle$ in Eq.\eqref{Wick-mom} which reduces to
\begin{equation}
\langle v^2 \rangle	:=\langle v^2 \rangle^{(0)} + \frac{1}{\alpha}\langle v^2 \rangle^{(1)} = \Sigma_{22}^{(0)}  + \frac{1}{\alpha}\Sigma_{22}^{(1)} ,
\end{equation}
and leads to 
\begin{equation}
	T_s = T_b -\frac{1}{\alpha}  \frac{\dot{T_b}}{2\gamma}.
\end{equation}
The temperature of the system in oscillating state begins to deviate from the bath temperature$~T_b$ as$~\alpha$ decreases from infinity, where the deviation depends on the rate at which bath temperature changes. Of course the temperature is expected in general to depend on the potential when$~x$ and$~v$ degrees are correlated as is also evident from Eq.\eqref{Wick-mom}. We find though that $T_s$ is oblivious to the potential to sub-leading order. 

The correlation function can be explicitly obtained from Eq.\eqref{Wick-mom} and reads as
\begin{equation}\label{c-Or1}
	C= \frac{1}{\alpha}\frac{1 }{\sqrt{\Sigma_{11}^{(0)}\Sigma_{22}^{(0)}}} \left( 1+ 6a_0^{(0)}\Sigma_{11}^{(0)2} \right) \langle xv \rangle^{(1)},
\end{equation}
where the sub-leading correction of the moment$~\langle xv \rangle$ is given as
\begin{equation}
	\langle xv \rangle^{(1)} = \Sigma_{12}^{(1)}-3 \left(\Sigma_{11}^{(0)}\right)^2 \left(
	4 a_0^{(0)} \Sigma_{12}^{(1)} 
	+ a_1^{(1)}  \Sigma_{22}^{(0)} \right).
\end{equation}
We find that position and velocity variables become correlated in the oscillating state at sub-leading order. Note that the correlations depend on$~\Sigma_{12}^{(1)}$ which is proportional to the difference between$~T_b$ and$~c^{(0)} k$ or equivalently the difference between kinetic and harmonic potential energies at leading order. In absence of anharmonic perturbation for which$~a_0^{(0)}$ and$~a_1^{(1)}$ vanish,  the correlation function quantifies a violation of equipartition of energy in the oscillating state. The sub-leading correction in the general case can be expressed as
\begin{equation}\label{C-h-ah}
\langle xv \rangle^{(1)} =\frac{1}{\gamma} \left[T_b -c^{(0)} k \left( 1- 12a_0^{(0)}\Sigma_{11}^{(0)2} \right) 
-  3 \lambda \Sigma_{11}^{(0)2} \right]~.
\end{equation} 
The explicit expression of the harmonic part is
\begin{equation}\label{xv-h-1}
	\Sigma_{12}^{(1)} = \frac{T_b}{\gamma} - \frac {\overline{(T_b/\gamma)}}{\overline{(k/\gamma)}}\frac{k}{\gamma},
\end{equation}
which is in general non-zero for driven cases unless we either fine tune$~T_b$ to$~c^{(0)} k$ or choose$~\gamma$ to be the only time-dependent parameter. Similarly, the explicit expression of the anharmonic part is
\begin{equation}\label{xv-an-1}
\langle xv \rangle^{(1)} - \Sigma_{12}^{(1)}
= -3\left( \frac {\overline{(T_b/\gamma)}}{\overline{(k/\gamma)}}\right)^2 \left[\frac{\lambda}{\gamma} - \frac {\overline{(\lambda/\gamma)}}{\overline{(k/\gamma)}}\frac{k}{\gamma} \right],
\end{equation} 
which is in general non-zero and vanishes either when both$~k$ and$~\lambda$ are constants or when driven with fine-tuning$~\lambda$ with respect to$~k$. It is evident that anharmonic drives can be made use to either enhance or reduce the corrections between$~x$ and$~v$. Furthermore, it is extremely interesting to note that the quantity which appears in the brackets apart from$~T_b$ in Eq.\eqref{C-h-ah} is related to the configuration temperature\,\cite{Casas2003}
\begin{equation}\label{conf-temp}
T_c := \frac{\langle U'^2 \rangle}{\langle U''\rangle} = k\langle x^2\rangle + 3 \lambda \Sigma_{11}^2~,
\end{equation}
where the primes on$~U =U(x,t)$ denote derivatives with respect to$~x$.
Thus we obtain the expression
\begin{equation}
	C= \frac{1}{\alpha \gamma \sqrt{T_b}} 
	\left[\frac{\overline{(k/\gamma)}} {\overline{(T_b/\gamma)}} +3\frac{\overline{(\lambda/\gamma)}} {\overline{(k/\gamma)}} \right]^{\frac{1}{2}} \lim\limits_{\alpha \to \infty} (T_s - T_c),
\end{equation}
which suggests that the correlation function may capture the difference between kinetic and configuration temperatures.

Since the quantities$~q_{hk}, \Phi$ and$~\Pi$ are of$~O(\alpha)$, we can expect that their expansion to$~O(1/\alpha)$ will contain the elements$~\Sigma_{ij}^{(2)}, a_i^{(2)}$ or$~\tilde{a}_i^{(2)}$. The expression for house-keeping heat flux obtained by substituting Eq.\eqref{Wick-mom} in Eq.\eqref{hk-heat} is given as
\begin{equation}
	q_{hk}= -\frac{1}{2}\dot{T_b}+ \frac{1}{\alpha}\gamma\left(\Sigma_{22}^{(2)} +  \langle v^2 \rangle^{(2)}_{ah} \right),
\end{equation}
where the anharmonic contribution to$~\langle v^2 \rangle$ at$~O(1/\alpha^2)$ can be written as
\begin{equation}
	\langle v^2 \rangle^{(2)}_{ah} = \frac{3c^{(0)}}{\gamma^2} \left[  c^{(0)} k \left(\lambda-4 c^{(0)}_0 c^{(0)} k \right) -\lambda \left(T_b -c^{(0)} k \right)\right],
\end{equation}
by straightforward substitutions. The two terms in the anharmonic contribution written within parenthesis are in fact the same expressions that have appeared in anharmonic and harmonic parts of the correlation function, respectively. Also note that at this order the flux depends on the second derivative of$~T_b$ contained in$~\Sigma_{22}^{(2)}$.

We see that even in the limit when$~\alpha \to \infty$ there is a non-zero heat flux required to sustain the oscillating state which is completely dictated by the rate with which bath temperature changes. Incidentally, the heat flux averaged over a time-period vanishes and almost insinuates the possibility that time-averaged probing may not distinguish oscillating state from equilibrium at the leading order. We can though easily rule out the possibility by considering, for instance, the time-averaged variance of house-keeping heat flux which is non-zero in oscillating state unlike in equilibrium. The quantity$~{(T_b-c^{(0)} k)}$ which measures the violation of equipartition of energy in oscillating state is non-zero in general even when time-averaged over a period and thus is yet another quantity to distinguish oscillating state from equilibrium even at leading order. 

The entropy flux$~\Phi= q_{hk}/T_b$ of course follows the heat flux and hence is non-zero at leading order. Its average over a period also vanishes at this order since heat flux is proportional to$~\dot{T_b}$. On the other hand, there is no entropy production at the leading order as is evident from Eq.\eqref{Wick-ent} where the terms of$~O(\alpha)$ and$~O(1)$ vanish. By direct substitutions we obtain the expression for$~\Pi$ at sub-leading order as
\begin{eqnarray}\label{Wick-ent-1}
	\Pi = \frac{1}{\alpha}
	\gamma \left[ \left( \frac{ \Sigma_{22}^{(1)}}{T_b}\right)^2
	+ \left(C^{(1)} \right)^2 \right],
\end{eqnarray}
where
\begin{equation}
 \left(C^{(1)} \right)^2 = \frac{(\Sigma_{12}^{(1)})^2}{T_b}\left( \frac{1}{\Sigma_{11}^{(0)}}- 12a_0^{(0)}\Sigma_{11}^{(0)} \right) -
 6a_1^{(1)}\Sigma_{11}^{(0)} \Sigma_{12}^{(1)} ,
\end{equation}
It can easily be verified using Eq.\eqref{c-Or1} that$~C^{(1)}$ is in fact the correlation function, namely
\begin{equation}
	C^{(1)} = \lim\limits_{\alpha \to \infty} \left( \alpha C \right).
\end{equation}
It is of course expected that the entropy production is positive. We further find that it can be written as sum of two positive quantities each of which has simple physical interpretation. Essentially, the production rate can be  increased quadratically by either increasing rate of change of bath temperature or by enhancing the correlations between position and velocity variables.

The irreversible probability current$~J_v^{ir}$ does not vanish in the oscillating state unlike equilibrium. Thus the detailed balance condition does not hold in oscillating states. We can easily verify this in Eq.\eqref{Jirr-1} where$~O(\alpha)$ term vanishes ensuring a finite bound while$~O(1)$ term is non-zero. The expectation of current$~\langle J_v^{ir} \rangle$ is zero while the variance$~\langle (J_v^{ir})^2 \rangle$ is non-zero, as can be deduced from$~{(x,v) \to (-x,-v) }$ symmetry of$~P_{os}$. The variance of irreversible current is non-zero even in the limit$~{\alpha \to \infty}$ where there is no entropy production. The explicit expression for the variance in harmonic case to leading order is obtained as
\begin{equation}
\langle (J_v^{ir})^2 \rangle = \left(\frac{\gamma}{6\pi T_b c^{(0)}} \right)^2 
\left[ \frac{1}{4} c^{(0)} \dot{T}_b^2 + T_b \left(T_b -c^{(0)} k \right)^2 \right].
\end{equation}
The variance is found to be sum of two positive quantities each quantifying the violation of detailed balance. The first one vanishes only when bath temperature is constant and the second one when position and velocity variables are uncorrelated. The sub-leading expressions including anharmonic perturbations can also be calculated straightforwardly though are cumbersome to express. 

The observables that we come across in equilibrium such as energy and entropy also dependent on$~\gamma$ in oscillating state. The leading and sub-leading terms of energy 
\begin{equation}
	E =  E^{(0)}  + \frac{1}{\alpha}E^{(1)},
\end{equation} 
can be easily read off from Eq.\eqref{ener-exp}. The leading term of energy$~E^{(0)}$ apart from$~T_b$ depends on$~\gamma$ through$~\langle x^2 \rangle^{(0)}$. The sub-leading term of energy reduces to
\begin{equation}
	E^{(1)}= -\frac{\dot{T_b}}{4\gamma} + \frac{1}{2} k\langle x^2\rangle^{(1)} + \frac{3}{2} \lambda \Sigma_{11}^{(0)}\Sigma_{11}^{(1)}~,
\end{equation} 
and is non-zero in oscillating state unlike equilibrium. This term clearly depends on$~\gamma$, the rate of change of the bath temperature and an accumulated violation of equipartition due to$~\Sigma_{11}^{(1)}$. The expression further contains the sub-leading correction$~\langle x^2 \rangle^{(1)}$ of the moment$~\langle x^2 \rangle$ which can be obtained from Eq.\eqref{Wick-mom} and reads
\begin{equation}\label{x2-1}
	\langle x^2 \rangle^{(1)} =  \Sigma_{11}^{(1)} -2\Sigma_{11}^{(0)2}\left(\tilde{a}_0^{(1)} +6 a_0^{(1)} \Sigma_{11}^{(0)} 
+ 18 a_0^{(0)} \Sigma_{11}^{(1)} \right),
\end{equation}
which can be recast by explicit substitutions and cumbersome but straightforward algebraic manipulations as
\begin{equation}\label{x2mom-1}
\langle x^2 \rangle^{(1)} (t) = c' + 2 \int_{0}^{t}ds \langle xv \rangle^{(1)}(s)~,
\end{equation} 
where the constant
\begin{equation}
c' =c^{(1)}\left(1-36 c^{(0)2}c^{(0)}_0 \right) - 2 c^{(0)2} \tilde{c}^{(1)}_0- 12 c^{(0)3} c^{(1)}_0.
\end{equation} 
The expression for$~\langle x^2 \rangle^{(1)}$ as given in Eq.\eqref{x2mom-1} allows it to be interpreted in terms of accumulated correlations of position and velocity variables using  Eqs.\eqref{xv-h-1} and~\eqref{xv-an-1}. 

Similarly, the leading and sub-leading terms of entropy 
\begin{equation}
	S =  S^{(0)}  + \frac{1}{\alpha}S^{(1)},
\end{equation} 
can be read off from Eq.\eqref{entr-exp}. Though the limiting expressions obtained using Eq.\eqref{Wick-mom} are not particularly revealing, the sub-leading term of entropy can be recast elegantly as 
\begin{equation}\label{s1-term}
	S^{(1)} = \frac{1}{2} \left[\frac{\langle v^2 \rangle^{(1)}}{\langle v^2 \rangle^{(0)}}  +  \frac{\langle x^2 \rangle^{(1)}}{\langle x^2 \rangle^{(0)}} \right],
\end{equation}
where its dependence on$~\gamma, T_b,\dot{T_b}$ and the accumulated correlations is more transparent. Though this expression is far from obvious, it is nevertheless straightforward to verify that Eq.\eqref{s1-term} is indeed the sub-leading term of Eq.\eqref{entr-exp}. The first term on the right in Eq.\eqref{s1-term} vanishes when bath temperature is kept constant while the second term vanishes when velocity and position are uncorrelated. 

To summarize, thermodynamic properties in oscillating state are drastically different from those in equilibrium even in large viscous regime. We can of course evaluate viscous effects beyond sub-leading order unlike in driven over-damped Langevin systems which is only valid to$~O(1/\alpha)$. 

\begin{figure*}[!htb]
	\centering
	\includegraphics[width=\linewidth]{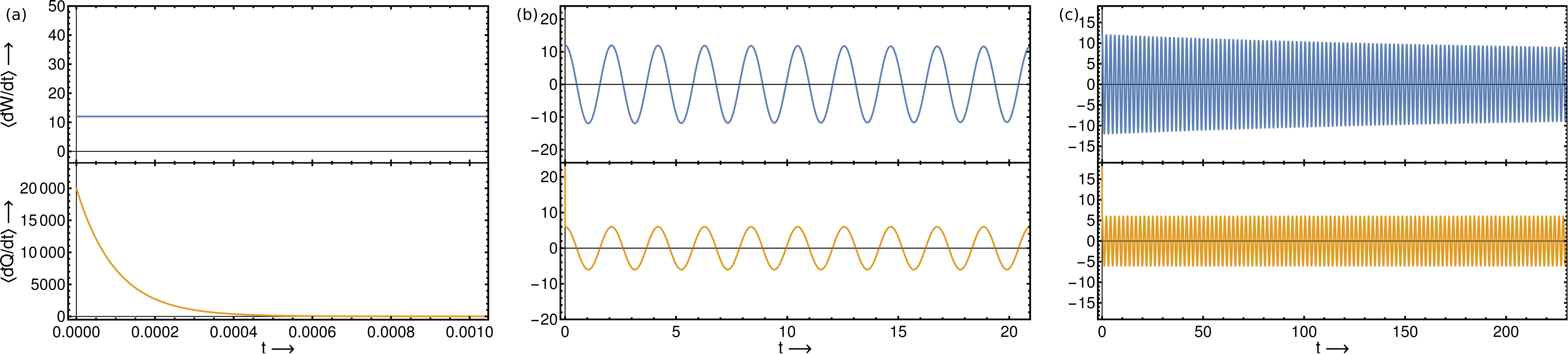}
	\caption{\label{dW-dQ-ti} Relaxation of rate of work done $\langle \dbar W/dt \rangle$ and rate of heat flux $\langle \dbar Q/dt \rangle$ vs time $t$ for $\gamma(t) = 3 + 2\cos(3 t)$, $T_b(t) = 6 + 4 \sin(3 t)$, harmonic potential strength $k(t) = 5 + 4 \sin(3 t)$, and $\alpha = 1000$}
\end{figure*}

It may be beneficial to digress briefly and compare with the over-damped case which is governed by Eq.\eqref{stoc-dyn} but without the inertial term, namely
\begin{equation}\label{od-eq}
	\gamma\dot{X}_t = f \left( X_t; k, \lambda \right) + \eta(t)~.
\end{equation}
For certain drives the values of the observables can significantly differ when evaluated even to$~O(1/\alpha)$ using over-damped instead of under-damped dynamics. For instance, consider the house-keeping heat flux in over-damped case\,\cite{Sekimoto1998, Sekimoto2010} given by
\begin{equation}
 q_{hk} \big|_\text{od} = \frac{1}{\gamma}\left( \left\langle \left( \frac{\partial U}{\partial x} \right)^2 \right\rangle_\text{od} - T_b \left\langle  \frac{\partial^2 U}{\partial x^2} \right\rangle_\text{od} \right),
\end{equation}
where the statistical averages$~\left\langle \cdots \right\rangle_\text{od}$ are with respect to the asymptotic probability distribution of the over-damped system. This asymptotic distribution in the limit$~\alpha \to \infty$ in fact coincides with the position variable marginal distribution of the under-damped oscillating state (see \hyperref[app1]{appendix}). Hence we obtain the expression
\begin{equation}
q_{hk}=q_{hk} \big|_\text{od}+
\frac{\dot{T}_b}{2} - \frac{1}{\alpha}\frac{1}{4\gamma} \frac{d}{dt} \left(\frac{\dot{T}_b}{\gamma} \right). 
\end{equation}
We find that when$~T_b$ is time dependent a naive over-damped analysis misses out both leading and sub-leading contributions that depend on$~\dot{T_b}$. The discrepancy due to over-damped approximation is of course expected and in fact studied in the context of B\"{u}ttiker-Landauer motor and refrigerator\,\cite{Benjamin2008,Sekimoto2010} and Heat Engines\,\cite{Ai2006,Arold2019}.

\section{Numerical Analysis}\label{num-ana}
In this section, we discuss the differences in relaxation timescales of different observables and then illustrate the efficacy of the perturbative scheme used by means of a numerical example. We restrict to harmonic potential for simplicity but the analysis can easily be extended to include anharmonic perturbations. The periodic functions$~\gamma, T_b$, and $k$ of course should be chosen such that the existence condition\,\eqref{cond-mu-g0} holds and the system thus is ensured to persist in an oscillating state. We consider the sample functions $\gamma(t) = 3 + 2 \cos(3 t)$, $T_b(t) = 6 + \sin(3 t)$ and $k(t) = 5 + 4 \cos(3 t)$ that have time-period$~T=2 \pi /3$. For this choice, $\overline{k/\gamma} \approx 2.24 >0$ and the system indeed relaxes asymptotically to an oscillating state in the large$~\alpha$ regime.

It should be reiterated that not all observables relax to their oscillating state values within the same timescale in the large viscous regime. For instance, the relaxation time of the moments$~\langle v^2 \rangle$ and$~\langle x v \rangle$ is of $O(\alpha^1)$, while that of$~\langle x^2 \rangle$ is $O(\alpha^0)$. Consequently, we expect the relaxation timescales of rates of heat flux$~\langle \dbar Q /dt \rangle$ and of work done $\langle \dbar W/dt \rangle$ to be drastically different from each other. In Fig.\,\ref{dW-dQ-ti}, we plot these two quantities in order to see the differences in their relaxation times from their respective initial values to their oscillating state values by choosing three different evolution timescales for a fixed$~\alpha=1000$. The first case is shown in Fig.\,\hyperref[dW-dQ-ti]{1(a)} where the dynamics of  both the quantities is followed from the start for roughly two-thousandth of a time period. We clearly observe that even within this short time interval, the heat flux rate has deviated significantly while the rate of work done hardly deviates from its initial value. The second case can be seen in Fig.\,\hyperref[dW-dQ-ti]{1(b)} which depicts the relaxation for almost 10 time periods from the start. Here it may appear that both quantities attained their oscillating state values due to their near periodic behavior. But in fact the rate of relaxation of $\langle \dbar W/dt \rangle$ is too slow to numerically realize during this time scale that the observable has not yet attained its oscillating state value. To notice the differences in relaxation times of the two quantities, we need to evolve even longer as shown in Fig.\,\hyperref[dW-dQ-ti]{1(c)} where evolution of the two quantities is plotted for approximately 100 time periods and from which it is evident that$~\langle \dbar W/dt \rangle$ is still evolving. Essentially, in the large viscous regime some observables can relax very slowly and caution is required to numerically assert their values in oscillating state. Note that all the plots and numerical calculations are done using \textit{Mathematica} and the error bars are less than the thickness of lines used in the plots.

\begin{figure}[t]
	\centering
	\includegraphics[width=\linewidth]{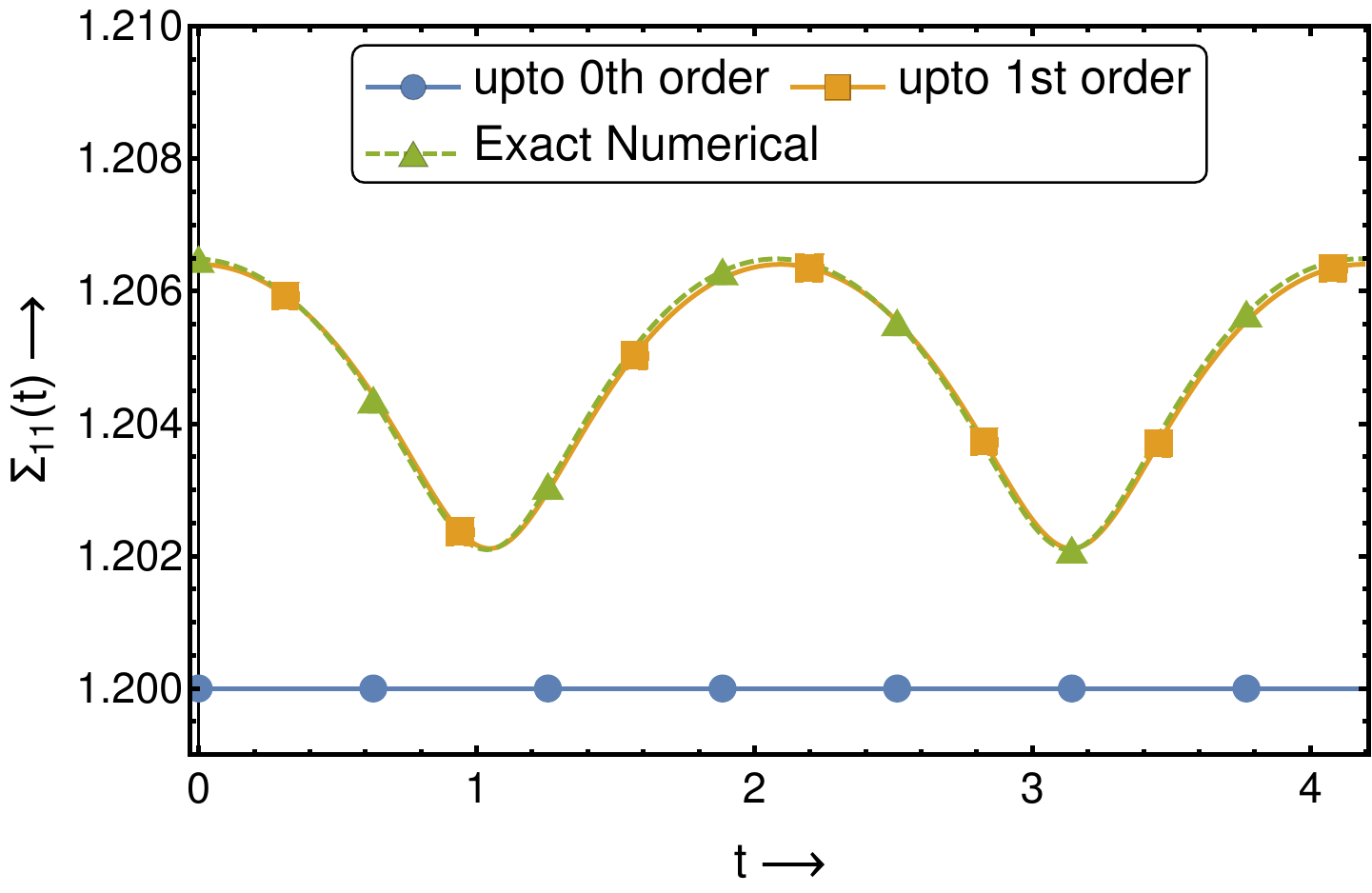}
	\caption{\label{v2-ti} Position auto-correlation $\Sigma_{11}(t)$ vs time $t$ for $\gamma(t) = 3 + 2\cos(3 t)$, $T_b(t) = 6 + 4 \sin(3 t)$, harmonic potential strength $k(t) = 5 + 4 \sin(3 t)$, and $\alpha = 100$}
\end{figure} 

We now compare numerical results with corresponding perturbative predictions. We choose the same sample functions for the drive parameters but a smaller value of$~{\alpha = 100}$ and compare the numerical and perturbative estimates of, say, the variance or auto-correlation of position$~\Sigma_{11}(t)$ in order to see the effectiveness of the perturbative scheme. The results are plotted in Fig.\,\ref{v2-ti} over a time domain of two periods so as to manifest the periodicity of oscillating state. We have seen from perturbative analysis that$~\Sigma_{11}^{(0)} = c^{(0)}$. This constant can easily be calculated by substituting the chosen sample functions in Eq.\eqref{c0-def} and is found to be$~c^{(0)}=1.2$. The first-order correction $\Sigma_{11}^{(1)}$ can also be determined by numerically integration using Eqs.\eqref{ite-mom},\eqref{sig-11-n} and\,\eqref{sig-11-hat}. We have also determined the asymptotic variance by numerically solving the exact dynamical equations\,\eqref{mom-2-as} for arbitrary initial values. In this case we have evolved the solution to over 200 time periods in order to ensure the periodicity to desired accuracy. We clearly observe in Fig.\,\ref{v2-ti} that the first-order approximation matches excellently with exact numerical solution. We have further confirmed though not shown here that the other two moments $\langle x v \rangle$ and $\langle v^2 \rangle$ also show similar agreement. Furthermore, even on adding an anharmonic potential, say, with the choice$~\lambda(t) = 0.01(2+\cos(3t))$ the perturbative results are in excellent agreement with the corresponding numerical solutions.

\section{Concluding remarks}\label{con-rem}
We have considered a class of periodically driven underdamped anharmonic Langevin systems parameterized by a set of $T$-periodic functions consisting of$~\gamma,T_b, k$ and$~\lambda$ corresponding to viscous, thermal, harmonic and anharmonic drives, respectively. Under certain conditions these systems can exist in oscillating states that we have explicitly explored for large viscous drives by a one-parameter extension of$~\gamma \to \gamma_{\alpha} = \alpha \gamma$. We have employed an appropriate large-$\alpha$ perturbative scheme to explicitly express the existence condition and determine the probability distribution of oscillating states to$~O(\lambda)$. The limit$~\alpha \to \infty$ is a singular limit and in turn demands cautiousness in obtaining the distribution of oscillating states when obtained by taking large time limit of the solutions of corresponding Fokker-Planck equation of the Langevin system.  In other words, the limits$~t \to \infty$ and$~\alpha \to \infty$ do not commute. This is also the reason for significant slowing down of relaxation time for position variable in these driven systems when disturbed from their oscillating state for large values of$~\alpha$. Nevertheless the relaxation of velocity variable and correlations between position and velocity variables are unaffected by this singular limit. 

The probability distribution obtained explicitly enabled us to determine various thermodynamic quantities to sub-leading order in$~1/\alpha$. We find that oscillating states are distinct and show measurable differences from equilibrium even in the limit$~\alpha \to \infty$ including violation of equipartition in harmonic case, presence of house-keeping heat and entropy fluxes and non-zero variance of irreversible probability current. Even more drastic differences emerge as we move away from this limit. The kinetic temperature$~T_s$ of the system in oscillating state begins to deviate from the bath temperature$~T_b$ as$~\alpha$ reduces from$~\infty$, where the deviation is controlled linearly by$~\dot{T_b}$ and being sub-leading correction inversely by$~\gamma$. The correlations between position and velocity begin to emerge in oscillating states proportional to the difference between kinetic and configuration temperatures. Entropy production begin to commence with a rate that has a quadratic dependence on both$~\dot{T_b}$ and the correlations. Sub-leading corrections of course add on to heat flux that are further sensitive to $~\ddot{T_b}$  and$~\dot{\gamma}$. In the oscillating state, entropy varies in time only due to heat exchange in the$~\alpha \to \infty$, while in case of finite$~\alpha$ the variation is also due to entropy production. We also found that sub-leading terms of energy and entropy  further depend on$~\dot{T_b}$ and accumulated correlations of position and velocity. To summarize, we have analyzed driven under-damped Langevin systems that perpetuate in oscillating states by exchanging heat flux with the bath and by generating entropy production, and explicitly studied the dependence of these states on various nonequilibrium quantities in the large viscous limit. We noticed in passing that neglecting the inertial term and restricting to over-damped approximation can lead to incorrect results for periodically driven Langevin systems. Finally, we have also numerically analyzed specific examples to accredit the employed perturbative scheme and to emphasize the differences in relaxation times of different observables.

In this work we have only considered quartic perturbations and analyzed the oscillating states to linear order in anharmonicity. The analysis can also be extended to other perturbations and even to higher order. In these cases it would be less cumbersome to first perturbatively solve the associated Hill equations and then evaluate observables by expressing them in terms of solutions of these Hill equations\,\cite{Awasthi2021} instead of directly finding the Taylor coefficients as we did here.  The perturbative scheme that we detailed in this work can also be easily extended to large viscous along with high frequency drives. The large$~\alpha$ expansion nevertheless has its limitation and is not suitable to evaluate observables in oscillating states when$~\alpha$ is small. It would be interesting to establish an appropriate perturbative scheme when$~\alpha \to 0$ and study the thermodynamics behavior for small viscous drives. We will explore in a future work the entire range of$~\alpha$ and investigate the dependence of observables in oscillating states beyond large viscous regime.

\appendix*
\section{Asymptotic distribution of overdamped Langevin equation}\label{app1}
The asymptotic distribution for the overdamped Langevin equation can easily be calculated in the limit$~{\alpha \rightarrow \infty}$ even when anharmonic perturbations are included, namely the force term in Eq.\eqref{od-eq} is taken to be$~f ( x; k, \lambda)=-kx-\lambda x^3$. In absence of anharmonic force, the asymptotic distribution is a periodic Gaussian distribution\,\cite{Awasthi2020} given by  
\begin{equation}
	P_{\text{od}}^{(0)}(x,t) = \frac{1}{\sqrt{2\pi \widetilde{X}_{2}\tm }}\;\; \mathrm{exp} \left[-\frac{x^{2}}{2\widetilde{X}_{2}\tm }\right]~,
\end{equation}
where$~\widetilde{X}_{2}$ is $T$-periodic asymptotic second moment$~\langle x^2 \rangle$. For the one-parameter extension $\gamma_\alpha = \alpha \gamma$ and fixed $T_b$, we obtain
\begin{equation}
	\lim_{\alpha \rightarrow \infty}\widetilde{X}_{2}(t) = \frac{\overline{(T_b/\gamma)}}{\overline{(k/\gamma)}} = c^{(0)}.
\end{equation}
When anharmonic force is also included, we can follow similar procedure as detailed in Ref.\cite{Awasthi2021}. We can essentially choose an ansatz for the asymptotic distribution to$~O(\lambda)$ of the form
\begin{equation}
	P_{\text{od}}(x,t) = P_{\text{od}}^{(0)} \left( 1 - (B^{(1)} - \langle B^{(1)} \rangle _{0} ) \right)~,
\end{equation}
where $\langle \cdots \rangle _{0} $ is average with respect to $P_{\text{od}}^{(0)}$ and
\begin{equation}
	B^{(1)} = \sum_{r=0} b_{r}(t) x^r~.
\end{equation}
On substituting the ansatz in the Fokker-Planck equation corresponding to Eq.\eqref{od-eq} and equating the coefficients of all independent monomials to zero, we can extract the dynamics of$~b_{r}$ and then determine the asymptotic distribution. We find that$~b_2$ and$~b_4$ are the only non-zero coefficients and 
\begin{eqnarray}
	\lim_{\alpha \rightarrow \infty} b_2(t) &=& 0, \nonumber \\
	\lim_{\alpha \rightarrow \infty} b_4(t) &=& \frac{1}{4} \frac{\overline{(\lambda/\gamma)}}{\overline{(k/\gamma )}} = c_{0}^{(0)}~.
\end{eqnarray}

\bibliography{RefShakul3a}
		
\end{document}